# A Wide Metallicity Range for Gyr-old Stars in the Nuclear Star Cluster

B. Thorsbro[1,2,3], R. Forsberg[3], G. Kordopatis[1], A. Mastrobuono-Battisti[4], R. P. Church[3], R. M. Rich[5], N. Ryde[3], M. Schultheis[1], and S. Nishiyama[6]

[1] Observatoire de la Côte d'Azur, CNRS UMR 7293, BP4229, Laboratoire Lagrange, F-06304 Nice Cedex 4, France; brian.thorsbro@fysik.lu.se
[2] Department of Astronomy, School of Science, The University of Tokyo, 7-3-1 Hongo, Bunkyo-ku, Tokyo 113-0033, Japan
[3] Lund Observatory, Division of Astrophysics, Department of Physics, Lund University, Box 43, SE-22100 Lund, Sweden
[4] GEPI, Observatoire de Paris, PSL Research University, CNRS, Place Jules Janssen, F-92190 Meudon, France
[5] Department of Physics and Astronomy, UCLA, 430 Portola Plaza, Box 951547, Los Angeles, CA 90095-1547, USA
[6] Miyagi University of Education, 149 Aramaki-Aza-Aoba, Aoba-ku, Sendai, Miyagi 980-0845, Japan

*Received 2023 September 7; revised 2023 October 26; accepted 2023 November 2; published 2023 November 21*

## Abstract

We report metallicities for three ∼Gyr-old stars in the Milky Way nuclear star cluster (NSC) using high-resolution near-infrared spectroscopy. We derive effective temperatures from a calibration with Sc line strength, which yields results in good agreement with other methods, and metallicities from spectral fits to Fe I lines. Our derived metallicities range from $-1.2 < [\text{Fe/H}] < +0.5$, a span of 1.7 dex. In addition we use isochrone projection to obtain masses of 1.6–4.3 $M_\odot$, and ages assuming single-star evolution. The oldest of these stars is 1.5 Gyr while the youngest and most metal-rich is only 100 Myr. The wide range in metallicity poses interesting questions concerning the chemical evolution and enrichment of the NSC and adds to the evidence for the presence of a young, metal-rich population in the NSC. We suggest that the candidate intermediate-age, metal-poor ([Fe/H] = −1.2) star may be best explained as a blue straggler from an underlying old population.

*Unified Astronomy Thesaurus concepts:* Stellar abundances (1577); Galactic center (565); Early-type stars (430)

## 1. Introduction

The Milky Way nuclear star cluster (NSC) surrounds the supermassive black hole (SMBH) Sagittarius A* at the very center of the Milky Way Galaxy. In general, many NSCs coexist with SMBHs, and can be found in more than 70% of galaxies with masses above $10^8$–$10^{10}\ M_\odot$ (see the review by Neumayer et al. 2020). The masses of SMBHs and NSCs have been shown to scale with the masses of their host galaxies, supporting a joint formation and evolution scenario (Kormendy & Ho 2013; Georgiev et al. 2016). The NSC hosts both very young <10 Myr and old, metal-rich populations.

Nishiyama et al. (2016, 2023) find candidates for a young/intermediate-age population, the subject of this work. For these candidate stars, there are no currently published metallicities measurements estimates. Determining their metallicity will enable an age determination to confirm whether they are indeed members of a young/intermediate-age population. The metallicity determination can also inform the chemical evolution and enrichment of the NSC, further potentially linking the stars to recent star bust events in the NSC or nearby environments (Nogueras-Lara et al. 2020, 2022; Schödel et al. 2023).

The star formation history of the NSC has been the subject of a number of investigations, beginning with Morris & Serabyn (1996), who proposed a continuous star formation history. More recently, Schödel et al. (2020) use the K-band luminosity function to propose that the bulk of the population is old (>10 Gyr) and metal-rich, with 15% of the mass being 3 Gyr old, and a few percent being 100 Myr or younger. Chen et al. (2023) challenge this picture using modeling of stellar metallicities; they find that 93% of the NSC is ∼5 Gyr old, with 7% of the mass having age 0.1–5 Gyr, with [M/H] = −1.1 dex.

Understanding the formation and evolution of the NSC is key both to gain a general understanding of the formation of the Milky Way and the NSCs of other disk galaxies (see the review by Neumayer et al. 2020). To explain the presence of an NSC in galaxies, two main formation channels have been proposed: (1) a cluster infall scenario where globular clusters (GCs) spiral into the center of the galaxy via dynamical friction. Due to mass segregation, the most massive stars will find themselves to the center most rapidly; nonetheless, the dynamical friction time for a GC is estimated to be 1 Gyr (Tremaine et al. 1975; Capuzzo-Dolcetta 1993; Antonini et al. 2012; Mastrobuono-Battisti & Capuzzo-Dolcetta 2012; Gnedin et al. 2014; Mastrobuono-Battisti et al. 2014; Perets & Mastrobuono-Battisti 2014; Antonini et al. 2015; Arca-Sedda et al. 2015; Guillard et al. 2016; Tsatsi et al. 2017; Abbate et al. 2018; Schiavi et al. 2021); and (2) in situ star formation (Loose et al. 1982; Levin & Beloborodov 2003; Milosavljević 2004; Nayakshin & Cuadra 2005; Paumard et al. 2006; Schinnerer et al. 2006, 2008; Hobbs & Nayakshin 2009; Mapelli et al. 2012; Mastrobuono-Battisti et al. 2019). It has been discussed that a combination of these two formation channels might be most plausible for the origin of the NSC in the Milky Way (Arca Sedda et al. 2020).

By considering the chemical composition of the stars, constraints can be put on the formation channels. Stars carry a chemical imprint of the molecular cloud from which they formed, acting as stellar fossils. Thus, detailed abundance trends allow chemical evolution models to constrain theories of the NSC's formation (Grieco et al. 2015; Matteucci et al. 2019; Thorsbro et al. 2020), with large predicted differences between the proposed formation scenarios. The abundance trends can also be compared with more well-studied Galactic populations, such as the thick disk, bulge, and bulge GCs.

**Table 1**
Compiled Data for the Observed Stars Discussed in This Paper

| Star ID | $M_{\rm bol}$ (mag) | $H$ (mag) | $K_s$ (mag) | R.A. (hh:mm:ss.sss) | Decl. (dd:hh:mm:ss) | $T_{\rm eff,N16}$ ±150 (K) | $T_{\rm eff,Sc}$ ±150 (K) | $\log g$ ±0.3 (dex) | $\xi_{\rm micro}$ ±0.2 (km s$^{-1}$) | [Fe/H] ±0.2 (dex) | Mass ($M_\odot$) | Age (Gyr) |
|---|---|---|---|---|---|---|---|---|---|---|---|---|
| 31 | −4.7 ± 0.2 | 12.1 | 10.2 | 17:45:39.616 | −29:00:44.46 | 3921 | 3833 | 0.50 | 2.4 | −0.6 | 3.1 ± 1.0 | $0.4^{-0.3}_{+0.5}$ |
| 36 | −4.3 ± 0.8 | 13.4 | 10.6 | 17:45:42.052 | −29:00:20.02 | 3527 | 3559 | 0.55 | 2.3 | 0.5 | 4.3 ± 1.0 | $0.1^{-0.08}_{+0.3}$ |
| 39 | −4.0 ± 0.3 | 13.2 | 11.2 | 17:45:39.880 | −28:59:56.49 | 4503 | 3972 | 0.80 | 2.2 | −1.2 | 1.6 ± 0.4 | $1.5^{-0.6}_{+3.2}$ |

**Note.** We assume solar abundances of $A$(Fe) = 7.45 (Grevesse et al. 2007). General uncertainties are listed for each of the stellar parameters; for mass and age we derive per-star uncertainties. The $T_{\rm eff}$ adopted and used in this paper is the $T_{\rm eff,Sc}$ determined using scandium lines; the $T_{\rm eff,N16}$ from Nishiyama et al. (2016) is listed for reference. The ID of the stars corresponds to the ID given by Nishiyama & Schödel (2013), Nishiyama et al. (2016), and those works are also the sources of the $M_{\rm bol}$, $H$, $K_s$, R.A., and decl. values.

The limited previous chemical studies of the NSC show a broad metallicity distribution, with progressively more metal-rich stars closer to the center (Ryde et al. 2016; Rich et al. 2017; Thorsbro et al. 2020). The first abundance trends were presented in Thorsbro et al. (2018), which revealed interesting star formation histories at supersolar metallicities. Furthermore, the first [α/Fe] trend of the old population of the NSC as presented in Thorsbro et al. (2020) demonstrates an intriguing trend for high [α/Fe] at higher metallicity.

In this work, we consider high-resolution infrared spectroscopy of three stars in the NSC with stellar parameters determined in Nishiyama et al. (2016). Using infrared observations allows us to penetrate through the high-extinction dust-covered regions toward the Galactic center. We determine the metallicities for the individual stars and provide an age estimation of the observed stars.

## 2. Observations

Three bright stars in the NSC, previously studied by Nishiyama & Schödel (2013) and Nishiyama et al. (2016), have been observed at medium/high-spectral resolution in the $K$ band using the NIRSPEC (McLean 2005; McLean et al. 2007) facility at Keck II under the program U095 (PI: Rich), as summarized in Table 1. The observations were obtained in the course of a larger program that produced the data for e.g., Rich et al. (2017).

We used the 0."432 × 12" slit and the NIRSPEC-7 filter, giving the resolving power of $R \sim 23{,}000$. The stars are observed with an ABBA scheme with a nod throw of 6" along the slit, in order to achieve proper background and dark subtraction. Five spectral orders are recorded, covering the wavelength range of 21000–24000 Å. The wavelength coverage is not complete; there are gaps between the orders. The stars are reduced with the NIRSPEC software `redspec` (Kim et al. 2015), and thereafter with IRAF (Tody 1993) to normalize the continuum, eliminate cosmic-ray hits, and correct for telluric lines. The latter has been done with a high-signal-to-noise spectrum of the rapidly rotating O6.5V star HIP89584. More details about the data reduction are given in Rich et al. (2017).

## 3. Analysis

Our analysis of the observed stars is described in the following subsections. First, the stellar parameters of effective temperature, surface gravity, and microturbulence are derived. This is followed by the line lists used for the spectral analysis and determination of metallicity. Finally, we describe how we derive our ages.

### 3.1. Effective Temperature, $T_{eff}$

The effective temperature, $T_{\rm eff}$, has already been determined by Nishiyama et al. (2016), and is included in Table 1 under the column $T_{\rm eff,N16}$. We also employ the method developed by Thorsbro et al. (2020) to derive $T_{\rm eff}$ using the high temperature sensitivity of the scandium lines arising from the 3d$^2$4s–3d4s4p transition (Thorsbro et al. 2018; Thorsbro 2020). The effective temperature determined using this method is included in Table 1 under the column $T_{\rm eff,Sc}$. The uncertainty of the effective temperatures for both methods is on the order of 150 K.

In cool stars blending of CN molecular lines has to be carefully accounted for when doing spectral analysis. To ensure that the scandium lines used for the temperature determination are not affected by CN blends, we model the spectra and vary the abundances of carbon and nitrogen with ±0.2 dex, shown in Figure 1. The line lists used in this modeling are the same as those used for our later analysis and are detailed in Section 3.4.

For stars 31 and 36 there is good agreement within 100 K in the temperature determinations for the two methods. For star 39 the Sc temperature is 500 K cooler. We examine the spectral lines of the three stars, as shown in Figure 1. Neutral scandium becomes a minority species around 4000 K. We model star 39 with an effective temperature of 4500 K as originally determined by Nishiyama et al. (2016) and find that the scandium features have almost fully vanished in the model spectrum, also shown in Figure 1 in the lower panels. However, in the observed spectrum we see scandium features for star 39, so we lean toward the temperature determination using the empirical scandium relation as more accurate. For our further analysis, we use the scandium-based temperature determinations.

### 3.2. Surface Gravity, $\log g$

For the surface gravity, we explore the use of photometry to determine the effect. In Nishiyama et al. (2016) the bolometric magnitudes are determined and the masses of the stars are estimated to be between 2 and 6 $M_\odot$. Together with the fundamental relation

$$\log\left(\frac{g}{g_\odot}\right) = \log\left(\frac{M}{M_\odot}\right) + 4\log\left(\frac{T_{\rm eff}}{T_{\rm eff,\odot}}\right) - 0.4(M_{{\rm Bol},\odot} - M_{{\rm Bol},*}), \quad (1)$$

this gives a range of surface gravity values for each star. For stars 31 and 36 the ranges are similar, being (0.25,0.73) and (0.28,0.76) respectively. For star 39 the range is (0.59,1.07). We adopt a log $g$ value in the middle of the range with a sufficiently large uncertainty of 0.3 dex. The adopted values are shown in Table 1.

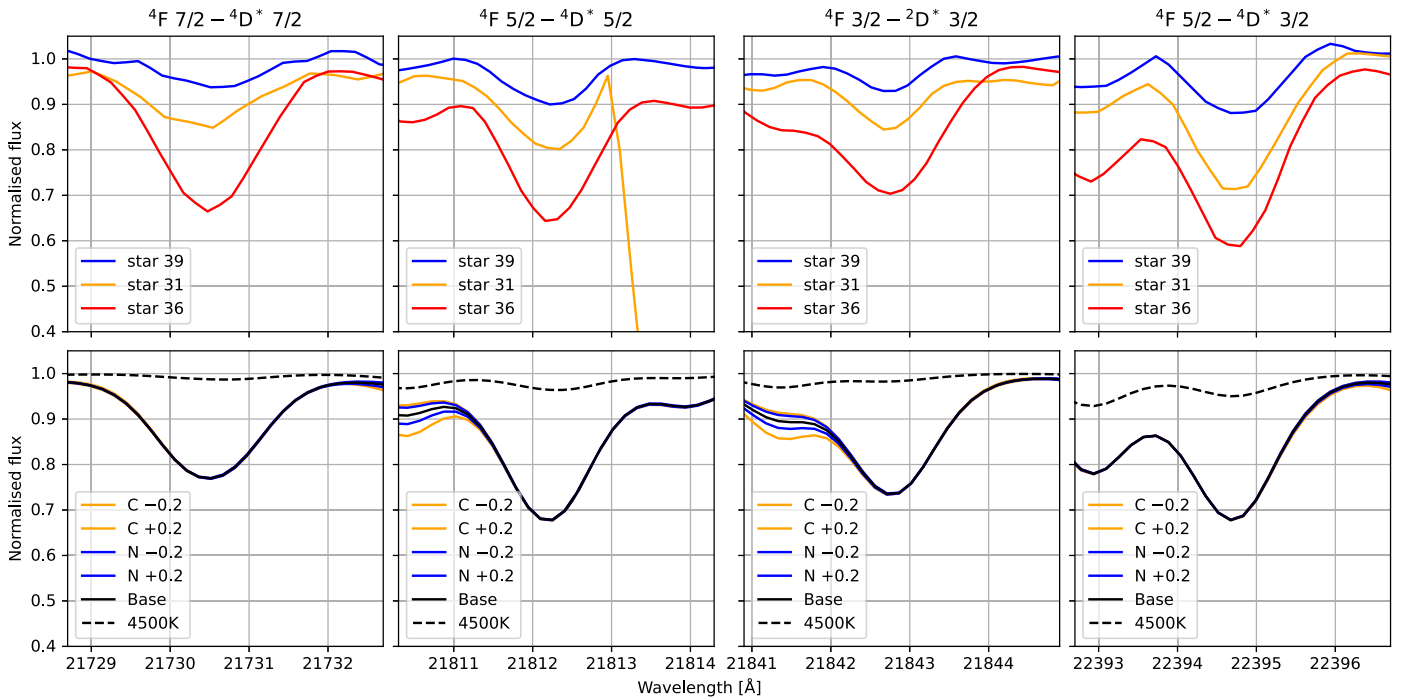

**Figure 1.** Four fine-structure energy level transitions in neutral scandium 3d²4s–3d4s4p. In the top panel: the top line is the observed spectra of star 39; middle and lower lines are star 31 and 36 respectively, ordered by decreasing $T_{\rm eff}$ from top to bottom. In the bottom panel: model spectra showing the sensitivity of the scandium lines to varying the carbon and nitrogen abundances, showing that the scandium lines are not affected by blending of CN molecular line features. The dashed line is a model of star 39 using an effective temperature of 4500 K and illustrates the vanishing scandium features as scandium ionizes, thus showing that 4500 K is probably too high for the temperature estimation.

### 3.3. Microturbulence, $\xi_{micro}$

The microturbulence, $\xi_{\rm micro}$, which takes into account the small scale, nonthermal motions in the stellar atmospheres, is important for saturated lines, influencing their line strengths. We estimate this parameter from an empirical relation with surface gravity based on a detailed analysis of spectra of five red giant stars ($0.5 < \log g < 2.5$) by Smith et al. (2013). The high-quality spectra analyzed by Smith et al. (2013) are from the archives of the Kitt Peak National Observatory Fourier transform spectrometer and their result are shown in Figure 2 together with our fit. The estimated uncertainty of 0.3 dex for surface gravity propagates into an uncertainty of 0.2 km s$^{-1}$ for $\xi_{\rm micro}$. In this study the uncertainty of $\xi_{\rm micro}$ does not dominate the uncertainty of the metallicity determination as we are using weak lines low on the curve of growth.

### 3.4. Line List

In order to synthesize the spectra, an accurate list of atomic and molecular energy level transitions is required. In the list of atomic energy level transitions provided by Thorsbro et al. (2017) wavelengths and line strengths (astrophysical log $gf$-values) are updated using the solar center intensity atlas (Livingston & Wallace 1991).

Because molecular lines are strong features in our spectra, we include molecular lines in the line list. For CN—which is the most dominant molecule apart from CO, whose lines dominate in the 2.3 $\mu$m bandhead region—we include the list of Sneden et al. (2016). The CO line data are from Goorvitch (1994). At the shorter wavelengths of our spectral region SiO, $H_2O$, and OH are important. Line lists for these molecules are included in our line list from, respectively, Langhoff & Bauschlicher (1993), Barber et al. (2006), and Brooke et al. (2016).

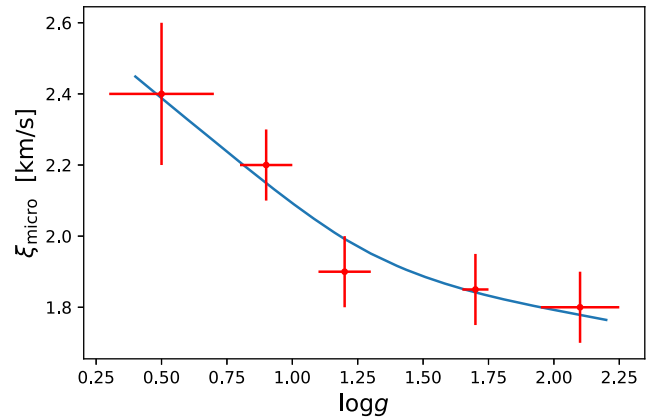

**Figure 2.** Empirical relation for $\xi_{\rm micro}$ vs. $\log g$, with the red crosses marking results from careful analysis of Smith et al. (2013) and the blue line our fit.

To identify clean spectral lines suitable for measurement, we have previously examined the synthetic model spectra of cool M giants to identify iron lines that are not blended with molecular features. For more details, we refer to Thorsbro et al. (2020). For the determination of metallicity, the iron lines shown in Table 2 have been identified as good lines with little or no blending.

### 3.5. Metallicity

In order to determine metallicities, we synthesize spectra and compare them to the observed spectra. We have chosen to use the spectral synthesis code Spectroscopy Made Easy (SME; Valenti & Piskunov 1996, 2012), which interpolates on a grid of one-dimensional MARCS atmosphere models (Gustafsson et al. 2008). These are hydrostatic model atmospheres in spherical geometry, computed assuming local thermodynamic equilibrium, chemical equilibrium, homogeneity, and conservation of the total flux (radiative plus convective, with the convective flux computed using a mixing-length recipe). The SME code has the advantage that it includes a flexible $\chi^2$ minimization tool to find the solution that best fits an observed spectrum in a prespecified spectral window. It also has a powerful continuum normalization routine. In cool star spectra, extra care is needed to normalize the spectrum in the region of a spectral line under consideration. The ubiquitous molecules and unavoidable residuals from the telluric line division necessitate careful analysis line by line. The uncertainty of our determination of metallicity is 0.1 dex from the method alone, but together with the high log $g$ uncertainty, which propagates into the metallicity uncertainty due to the degeneracy between the two parameters, we find an uncertainty of 0.2 dex for the metallicity.

**Table 2**
Lines Used for Abundance Determinations

| Wavelength in Air [Å] | $E_{exc}$ [eV] | log $gf$ |
| --- | --- | --- |
| Fe I | | |
| 21178.211 | 3.0173 | −4.201 |
| 21238.509 | 4.9557 | −1.281 |
| 21284.348 | 3.0714 | −4.414 |
| 21851.422 | 3.6417 | −3.506 |
| 22380.835 | 5.0331 | −0.409 |
| 22385.143 | 5.3205 | −1.536 |
| 22392.915 | 5.0996 | −1.207 |
| 22419.976 | 6.2182 | −0.226 |

### 3.6. Age and Mass Determination

The age and initial mass determinations were obtained using an improved version of the method presented in Kordopatis et al. (2023). In a nutshell, this method projects on a set of isochrones any combination of observed quantities such as effective temperature, surface gravity, metallicity and magnitudes, taking into account their associated uncertainties. From this projection, the expected theoretical stellar parameters are obtained.

In this work, the improvement consists on the computation of an age probability distribution function for each star in order to determine asymmetric uncertainties. The projection is performed by considering a flat prior between the age and the metallicity of the stars, and adopting the PARSEC isochrones (Bressan et al. 2012), which cover metallicities from [Fe/H] = −2.2 to +0.7 (with a step of 0.05 dex), and ages from 0.1 to 13.5 Gyr (with a logarithmic step). The observed parameters taken into account are $T_{eff}$, [Fe/H], and the absolute 2MASS $K_s$ magnitude. To obtain the latter, we assumed that the stars are located at the Galactic center distance of $8.275 \pm 0.042$ kpc (GRAVITY Collaboration et al. 2021) and we adopt the $A_{K_s}$ extinctions of Nishiyama et al. (2016). Uncertainties on the apparent magnitudes were assumed to be 0.05 mag and on $A_{K_s}$ of 0.1 mag. We note that additional projections were investigated, taking into account MIST isochrones (Dotter 2016), or different parameter combinations such as ($T_{eff}$, log $g$, [Fe/H]), ($T_{eff}$, [Fe/H], log $g$, $K_s$) or ($T_{eff}$, log $g$, [Fe/H], $K_s$, and $M_{bol}$). All of them gave similar results, within the uncertainties.

Overall, we find that all of the considered stars are relatively young, with two of them, younger than 0.5 Gyr. From the most metal-poor target to the most metal-rich, the derived ages are $1.5^{-0.6}_{+3.2}$, $0.4^{-0.3}_{+0.5}$, and $0.1^{-0.08}_{+0.3}$ Gyr, respectively. Their associated masses are $1.6 \pm 0.4$, $3.1 \pm 1.0$, and $4.3 \pm 1.0\,M_\odot$.

Figure 3 shows the location of the observations in the $T_{eff}$–$K_s$ diagrams with the measured values as black points and their associated error bars. The shaded colored areas correspond to the range covered by the isochrone points that fall within the age and metallicity uncertainties, and the dashed line corresponds to the isochrone of the reported age. To clarify the uncertainty of the age determinations, we show the posterior distribution function of the ages of the stars in Figure 4, with the median being the reported age.

### 4. Results/Discussion

Our derived ages imply that two of the three stars are apparently young, between 100 and 400 Myr, and one star has an intermediate age of 1.5 Gyr, broadly confirming the isochrone-derived ages of Nishiyama et al. (2016). However, the metallicities of all three young stars are separated by nearly 2 orders of magnitude, ranging from [M/H] = −1.2 dex to [M/H] = +0.50 dex. This separation is too large to originate from measurement errors and might indicate a real and interesting wide metallicity spread in the NSC young population. In accordance with galaxy and stellar evolution theory, these results point at the stars originating from chemically discrete molecular clouds. This can either be explained by that the NSC is not well mixed and has regions of lower-metallicity gas and regions of higher-metallicity gas, or that some of the stars have migrated into the NSC. The migration scenario, however, is challenged by the very young ages of the stars; it is highly unlikely that they could have traveled any vast distances in such a short amount of time, and the typical GC dynamical friction time is estimated to be 1 Gyr (Guillard et al. 2016). In a third scenario, the stars arise in a recent self-enriching burst of star formation.

### 4.1. A Blue Straggler

The metal-poor star with [M/H] = −1.2 dex, star 39, has a very young age for such a low metallicity. However, the derived age assumes that the star is, and has always been, single. A possible origin for star 39 could be that the star is a blue straggler (Sandage 1953). These are stars that have acquired mass from another star, either by mass transfer from a companion in a bound binary system or by physical collision and merger in a stellar cluster. The mass-gaining star in such a system then appears, when interpreted through the lens of single-star models, to be much younger than it actually is.

As a simple test of this scenario we have used the rapid stellar population synthesis code binary-star evolution (BSE) algorithm (Hurley et al. 2002) to compute the evolution of a binary of metallicity[7] [Fe/H] = −1.3 dex containing stars of masses 0.9 and $0.8\,M_\odot$ in a circular orbit initially of separation $6\,R_\odot$. In this rather close binary tidal circularization brings the two stars together, and the primary fills its Roche lobe after around 8.3 Gyr; a phase of stable mass transfer is followed by the formation of a contact binary as the accreting secondary star swells up and the binary merges at an age of 9.0 Gyr with negligible mass loss. The resulting star evolves thereafter as a single star and passes through the position of star 39 in the H-R diagram on the AGB at an age of 10.2 Gyr, with a mass of 1.64 $M_\odot$. We do not present this as a definitive model for the formation of star 39, but rather as an example to show that the star's observed properties are consistent with a very substantial age. We note that this binary would be dynamically hard even at separations as small as 0.3 pc from Sgr A$^*$ (Rose et al. 2020) and as such would be expected to survive in situ despite the harsh dynamical environment in the central parts of the Milky Way. While further exploration of this hypothesis would be difficult, increasing the number of young/intermediate-age stars with metallicities will shed light on how rare this star might be.

---

[7] The BSE algorithm empirically computes models most reliably when confined to metallicities that match the input tracks, in this case with a metal mass fraction of $Z = 0.001$.

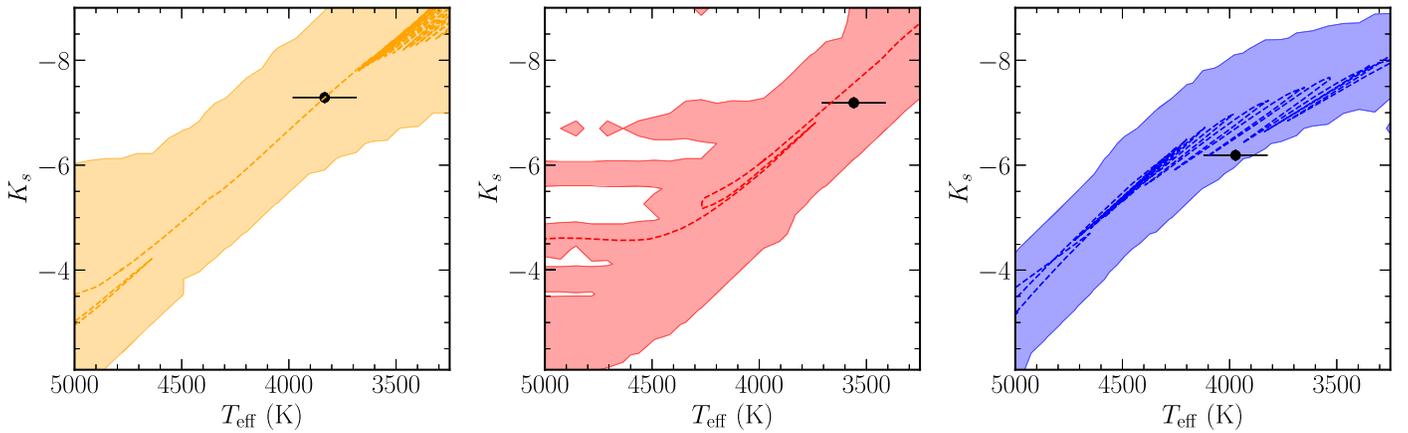

**Figure 3.** Measured parameters and uncertainties for the three considered stars (star 31 in orange, star 36 in red, star 39 in blue). The color-filled areas contain the isochrone points that are within the age and metallicity range of each star. The dashed line corresponds to the isochrone with the reported age.

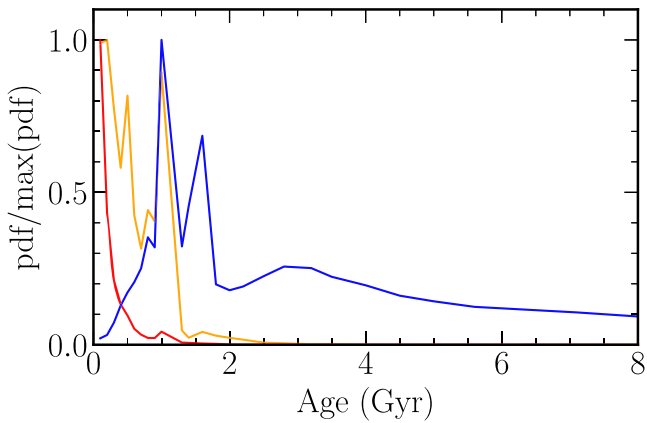

**Figure 4.** Posterior distribution function (PDF) of the ages of the stars (normalized to the peak) considered for this study. The plot has been truncated at 8 Gyr to be able to better see the PDF shapes at the younger ages. The adopted ages, reported in Table 1, correspond to the median values of the PDF.

## 5. Conclusions

We report new ages, masses, and metallicities for three candidate young stars from the sample of Nishiyama et al. (2016). We use high-resolution infrared spectroscopy to self-consistently measure metallicity and effective temperature, the latter using a new calibration based on Sc lines. We find a surprisingly wide range in metallicity for these stars, given that they all appear younger than ∼1.5 Gyr.

Although the sample is way too small to draw any inference on an age–metallicity relationship, it is interesting that our two youngest stars, 31 and 36, are [Fe/H] = −0.6 and +0.5, respectively, more metal-rich than the −1.1 dex advanced by Chen et al. (2023) for their "young/intermediate-age" component. The suggestion of an age–metallicity relationship is only that, with any confirmation requiring a significantly larger sample.

This work shows that infrared spectroscopy can be used to examine the chemical evolution of an interesting population in the evolution of the NSC: stars in the young to intermediate age range. As larger samples become available, it will be possible to explore the chemistry and dynamics of this interesting population in greater depth. The striking range in metallicity indicates a population of unexpected complexity.

More observations will be required to determine whether we are seeing evidence of star formation in unmixed clouds, migration of young stars into the NSC, the signature of a recent self-enriching burst of star formation, or if the low-metallicity star is a blue straggler. Finding such diverse metallicities is surprising, given the few parsec volume under consideration. Any further conclusions will require a significantly larger sample. We conclude that the population of the NSC in the age range between 100 and 1500 Myr is interesting and worthy of further study.

### Acknowledgments

B.T. acknowledges the financial support from the Wenner-Gren Foundation (WGF2022-0041). R.F. acknowledges support from the Royal Physiographic Society in Lund through the Stiftelse Walter Gyllenbergs fond and Märta och Erik Holmbergs donation as well as support from the Göran Gustafsson Foundation for Research in Natural Sciences and Medicine. N.R. acknowledges support from the Swedish Research Council, VR (project No. 621-2014-5640) and Kungl. Fysiografiska Sällskapet i Lund (Stiftelsen Walter Gyllenbergs fond and Märta och Erik Holmbergs donation). M. S. acknowledges the Programme National de Cosmologie et Galaxies (PNCG) of CNRS/INSU, France, for financial support. R.M.R. acknowledges financial support from his late father Jay Baum Rich. A.M.B. acknowledges funding from the European Union's Horizon 2020 research and innovation program under the Marie Skłodowska-Curie grant agreement No. 895174. R.C. acknowledges support from the Swedish Research Council, VR (project No. 2017-04217).

The data presented herein were obtained at the W. M. Keck Observatory, which is operated as a scientific partnership among the California Institute of Technology, the University of California, and the National Aeronautics and Space Administration. The Observatory was made possible by the generous financial support of the W. M. Keck Foundation. The authors wish to recognize and acknowledge the very significant cultural role and reverence that the summit of Maunakea has always had within the indigenous Hawaiian community. We are most fortunate to have the opportunity to conduct observations from this mountain.

*Facility:* KECK:II (NIRSPEC).

*Software:* SME (Valenti & Piskunov 1996, 2012), REDSPEC (Kim et al. 2015), IRAF (Tody 1993), BSYN & EQWIDTH (Gustafsson et al. 2008).

### ORCID iDs

B. Thorsbro ● https://orcid.org/0000-0002-5633-4400
R. Forsberg ● https://orcid.org/0000-0001-6079-8630
G. Kordopatis ● https://orcid.org/0000-0002-9035-3920
A. Mastrobuono-Battisti ● https://orcid.org/0000-0002-2386-9142
R. P. Church ● https://orcid.org/0000-0001-9204-0779
R. M. Rich ● https://orcid.org/0000-0003-0427-8387
N. Ryde ● https://orcid.org/0000-0001-6294-3790
M. Schultheis ● https://orcid.org/0000-0002-6590-1657
S. Nishiyama ● https://orcid.org/0000-0002-9440-7172